\documentclass[12pt]{article}
\usepackage{amsmath, amssymb, cases}
\newcommand{\ud}{\mathrm{d}}
\newcommand{\hl}{{L^2(\Omega,\mathbb{R}^3)}}
\newcommand{\hs}{{L^2(\Omega)}}
\newcommand{\ran}{ \textrm{Im }}
\newcommand{\Ran}{ \textrm{Ran }}
\newcommand{\krn}{ \textrm{Ker }}
\newcommand{\grad}{ \textrm{grad\,}}
\newcommand{\rot}{ \textrm{rot\,}}
\newcommand{\dive}{\textrm{div\,}}
\newcommand{\hc}{\mathcal{H}}
\newcommand{\E}{\textbf{E}}
\newcommand{\B}{\textbf{B}}
\newcommand{\V}{\textbf{V}}
\newcommand{\U}{\textbf{U}}
\newcommand{\n}[1]{\textbf{#1}}
\newcommand{\p}{\partial}

\def\tr{\operatorname{tr}}
\def\Tr{\operatorname{Tr}}
\def\const{{\rm const}\,}

\def\e{{\rm e}}
\def\iu{{\rm i}}

\def\o1{{\mathrm{o}(1)}}
\title{The heat kernel expansion for the electromagnetic field in a cavity}
\author{\hspace{-.2 cm} F. Bernasconi${}^{(a)}$, G.M. Graf${}^{(b)}$,
D. Hasler${}^{(c)}$\\
\normalsize\it \hspace{-.5 cm}\\
\hspace{-.5 cm}\normalsize\it ${}^{(a)}$ Department of Mathematics,
ETH-Zentrum, 8092 Z\"urich, Switzerland\\
\normalsize\it \hspace{-.5 cm}${}^{(b)}$ Theoretische Physik,
ETH-H\"onggerberg,
8093 Z\"urich, Switzerland\\
\normalsize\it \hspace{-.5 cm}${}^{(c)}$ Department of
Mathematics, University of Copenhagen, 
2100 Copenhagen, Denmark}

\begin{document}
\maketitle
\vspace{0.4cm}
\begin{abstract}
We derive the first six coefficients of the heat kernel expansion for 
the electromagnetic field in a cavity by relating it to the expansion for 
the Laplace operator acting on forms. As an application we verify that
the electromagnetic Casimir energy is finite.

\end{abstract}
\section{Introduction}\label{INTRO}

The modes of an electromagnetic field in a cavity, taken together with their
unphysical, longitudinal counterparts, can be mapped onto the eigenstates 
of the Laplacian acting on the de Rham complex of a 3-manifold with boundary.
The electric and magnetic fields are thereby associated 
to forms of degree $p=1$ and $p=2$ respectively. In this correspondence 
transverse modes are associated with coexact, resp. exact forms, which 
permits to further map longitudinal modes to forms of degree $p=0$ and 
$p=3$. We will use this observation, which is explained in detail
in Sect.~\ref{proofs} below, to compute the first six
coefficients of the heat kernel expansion for the electromagnetic field 
in a cavity. The result is used to show in a simple way that the Casimir
energy in an arbitrary cavity with smooth boundaries is finite, a conclusion 
which has been reached previously \cite{BD2}. In an appendix the
derivation of the numerical coefficients of the expansion is presented.\\

We shall present a Hilbert space formulation of the classical 
Maxwell equations in a cavity $\Omega \subset \mathbb{R}^3$. In a
preliminary Hilbert space $L^2(\Omega,\mathbb{R}^3)$ we define 
the dense subspaces
\begin{eqnarray*}
 \mathcal{R}&=& \left \{ \V \in L^2(\Omega,\mathbb{R}^3)\mid  
\rot \V \in L^2(\Omega,\mathbb{R}^3) \right \}\;,\\
 \mathcal{R}_{0} &=& \left \{\V \in  \mathcal{R}\mid 
\langle \U,\rot \V \rangle =  \langle \rot \U, \V \rangle,\,
\forall \U \in \mathcal{R}\right \}
\end{eqnarray*} 
and the (closed) operator
\[
R=\rot \qquad \textrm{with domain}\quad \mathcal{D}(R)=\mathcal{R}_{0}\;.
\]
Its adjoint is then given as $R^{*}= \rot$ with 
$\mathcal{D}(R^{*})=\mathcal{R}$.
We remark that $R$, resp. $R^*$, is also the closure of $\rot$ 
defined on smooth vector fields $\V$ with boundary condition 
$\V_\parallel =0$ on the smooth boundary $\p \Omega$, resp. without 
boundary conditions.
This is what is meant when we later simply say that a differential 
operator is defined with (or without) a certain boundary condition.\\

The subspace
\begin{equation}\label{uno} 
\hc =
\left \{\V \in L^2(\Omega,\mathbb{R}^3)\mid\dive \V = 0 \right \}
\end{equation}
and its orthogonal complement in $L^2(\Omega,\mathbb{R}^3)$ are 
preserved by $R$ and, therefore, by $R^*$.
We will thus view them as operators on the physical Hilbert space $\hc$.
The Maxwell equations with boundary condition $\E_\parallel=0$ on the 
ideally conducting shell $\p \Omega$ can now be written as
\begin{equation}\label{maxwell} 
\iu \frac{\p}{\p t}
\left (\begin{array}{l}\E\\\B \end{array}\right)= 
M\left (\begin{array}{l}\E\\\B \end{array}\right)
\end{equation}
with
\[
M =\left (\begin{array}{cc} 0 & \iu R^*\\-\iu R & 0\end{array}\right )
= M^*\qquad \textrm{ on } \hc \oplus \hc\;,
\]
cf. \cite{L}. Since no boundary condition has been imposed on $\B$, we have 
$M(0,\B)=0$ for all $\B=\nabla\psi$ with $\psi$ harmonic, and hence
\begin{equation}\label{due}
\textrm{dim } \krn M = \infty \;.
\end{equation}
We shall compute the heat kernel trace
\[
\Tr'_{\hc \oplus \hc}(\e^{-tM^2}) =
\sideset{}{'}\sum_{k}\e^{-t\omega_k^2} \;,
\]
where $'$ means that the contributions of zero-modes, i.e., of eigenvalues 
$\omega_k = 0$ of $M$, have been omitted. This is necessary in view of 
(\ref{due}), but a more physical justification, tied to the application 
to the Casimir effect to be discussed later, is that zero-modes are not 
subject to quantization.\\

The square of $M$ is
\begin{equation}
M^2 = \left (\begin{array}{cc}R^*R & 0\\0 & RR^*\end{array}\right) = 
\left(\begin{array}{cc}-\Delta_{\E} & 0\\0 & -\Delta_{\B} \end{array}\right) 
\;,
\label{m2}
\end{equation}
where $\Delta_{\E}$, resp. $\Delta_{\B}$, is the Laplacian on 
$\hc$ with boundary conditions
\begin{equation}
\label{dueprimo}
\E_\parallel = 0\;,\qquad \textrm{resp.}\quad (\rot\B)_\parallel = 0\;.
\end{equation}
The operators $RR^*$ and $R^*R$ have the same spectrum, including 
multiplicity, except for zero-modes. Incidentally, we note that 
eigenfunctions $(\E,\B)$ corresponding to $\omega_k \ne 0$ satisfy 
$\B = -\iu\omega_k^{-1} \rot \E$ and hence, by Stokes' theorem, the 
boundary condition $\B_\perp = 0$, which we did not impose, but which is 
usually also associated 
with ideally conducting shells. Since $\p_t^2+M^2=(\iu\p_t-M)(-\iu\p_t-M)$,
each pair of non-zero eigenvalues of $R^*R$ and $RR^*$ corresponds to 
a single oscillator mode for (\ref{maxwell}). We will thus discuss the 
heat kernel asymptotics for
\begin{numcases}{
\frac{1}{2} \Tr'_{\hc \oplus \hc}(\e^{-tM^2})=}
  \Tr'_{\hc}\e^{t\Delta_{\E}}
\label{eq:trea}\\
  \Tr'_{\hc}\e^{t\Delta_{\B}}
\label{eq:treb}
\end{numcases}
\begin{equation}
\hskip 5cm \cong\sum_{n=0}^\infty a_n t^{\frac{n-3}{2}} \;,
\qquad ( t \downarrow 0)\;.
\label{eq:asexp}
\end{equation}
The coefficients $a_{n}$ are known, see e.g. \cite{BGKV}, for general 
operators of Laplace type. The direct application of such results is 
prevented by the divergence constraint in $\hc$, 
see~(\ref{uno}). In the next section we indicate how to 
remove it. First however we present the main result.\\

%
Let
\begin{equation*}
L_{ab}= (\nabla_{\n{e}_a}\n{e}_b,\n{n} )\;,\qquad (a,b=1,2)\;,
\end{equation*}
be the second fundamental form on the boundary $\p\Omega$ with inward 
normal $\n{n}$ and local orthonormal frame $\{\n{e}_1,\n{e}_2,\n{n}\}$.
We denote by $|\Omega|$ the volume of $\Omega$ and set
\[
f[\p\Omega] = \int_{\p\Omega} f(y)\ud y\;,
\]
where $\ud y$ is the (induced) Euclidean surface element on $\p\Omega$.
The corresponding Laplacian on $\p\Omega$ is denoted by $\nabla^2$.
\newtheorem{teo1}{Theorem}
\begin{teo1}\label{teo1}
Let $\Omega \subset \mathbb{R}^3$ be a compact, connected domain with 
smooth boundary $\p\Omega$ consisting of $n$ components of genera 
$g_1, g_2,\dots,g_n$. Then
\begin{eqnarray}
 a_0&=& 2(4\pi)^{-\frac{3}{2}}|\Omega|\;,\nonumber  \\
   a_1&=& 0  \;, \nonumber \\
   a_2&=& -\frac{4}{3}(4\pi)^{-\frac{3}{2}}(\tr L)[\p\Omega]\;,\nonumber   \\
  a_3&=& \frac{1}{64}(4\pi)^{-1}
\bigl(3(\tr L)^2-4\det L\bigr)[\p\Omega]-
\frac{1}{2}\sum_{i=1}^n(1+g_i) + 1\;, 
\label{eq:a3}\\
  a_4&=& \frac{16}{315}(4\pi)^{-\frac{3}{2}}
\bigl(2(\tr L)^3- 9\tr L\cdot\det L \bigr)[\p\Omega]\;,\nonumber \\
a_5&=& \frac{1}{122880}(4\pi)^{-1}
\bigl(2295(\tr L)^4 - 12440(\tr L)^2\det L + \nonumber \\
&&\qquad\qquad \qquad 
 +13424(\det L)^2 + 1200\tr L\cdot\nabla^2\tr L\bigr)[\p\Omega]\;.\nonumber 
\end{eqnarray}
\end{teo1}
We will give two partially independent proofs, based on (\ref{eq:trea}), 
resp. (\ref{eq:treb}). Their agreement is related to the index theorem, as it
may be seen from (\ref{m2}). A further, partial check of these 
coefficients has been made on the basis of general cylindrical domains 
and of the sphere, where a separation into TE and TM modes is possible. 

The coefficient $a_0$ was computed in \cite{W} (except for the factor $2$
replaced by $3$, as the divergence condition (\ref{uno}) was ignored),
$a_1,\,a_2$ in \cite{BB}. The coefficient $a_3$ is closely related to a result
of \cite{BD2}, as discussed in Sect.~\ref{casi}.\\

\section{Proofs}\label{proofs}

We consider the space of (square integrable) forms, 
$\Lambda(\Omega)= \bigoplus_{p=0}^n \Lambda_p(\Omega)$, on the manifold 
$\Omega$ with boundary, together with the exterior derivative 
$d_{p+1}: \Lambda_p(\Omega)\rightarrow \Lambda_{p+1}(\Omega)$ defined 
with relative boundary condition (\cite{G}, Sect.~2.7.1)
\[
\omega \big\rvert_{\p\Omega} = 0\;,
\]
as a form $\omega\rvert_{\p\Omega}\in \Lambda_p(\p\Omega)$.
For later use we recall that by the de Rahm theorem for manifolds 
with boundary (\cite{D} or \cite{G}, Thm.~2.7.3) we have
\begin{equation}
\label{cinque}
H_r^p(\Omega)\cong H_{n-p}(\Omega)\cong H_p (\Omega,\p\Omega)\;,
\end{equation} 
where $H_r^p(\Omega)= \krn d_{p+1}/ \ran d_p$ 
is the $p$-th relative cohomology group, $H_p(\Omega)$ is 
the $p$-th homology group, and $H_p(\Omega,\p\Omega)$ is 
the $p$-th relative homology group, i.e., the homology based on chains
mod $\p\Omega$. 

We shall henceforth restrict to $\Omega \subset \mathbb{R}^3$ as in 
Theorem~\ref{teo1}. Using either homology (\ref{cinque}), the dimension of 
$H_r^p(\Omega)$ is seen to be
\begin{equation}\label{sei}\begin{aligned}
0  &&\qquad\qquad&&(p=0)\;,\\
n-1 && &&(p=1)\;,\\
\sum_{i=1}^n g_i  && &&(p=2)\;,\\
1 && &&(p=3)\;.
\end{aligned}\end{equation}
These are also the dimensions of the spaces of harmonic $p$-forms.\\

The space $\Lambda(\Omega)= \bigoplus_{p=0}^3 \Lambda_p(\Omega)$ 
may be identified as
\begin{equation*}
\Lambda(\Omega) =  L^2(\Omega)\oplus
    L^2(\Omega,\mathbb{R}^3)\oplus L^2(\Omega,\mathbb{R}^3)\oplus
   L^2(\Omega)\ni (\phi,\E,\B,\psi)\;,
\end{equation*}
where $d : \Lambda(\Omega) \rightarrow  \Lambda(\Omega)$ acts as 
\begin{gather*}
d :L^2(\Omega)\underset{\text{grad}}{\longrightarrow}
    L^2(\Omega,\mathbb{R}^3)\underset{\text{rot}}{\longrightarrow} 
    L^2(\Omega,\mathbb{R}^3)\underset{\text{div}}{\longrightarrow}
    L^2(\Omega){\longrightarrow}0
\intertext{with boundary conditions 
$\phi = 0,\,\E_\parallel = 0,\, \B_\perp = 0$ on $\p\Omega$. Then}
d^* :0 \longleftarrow L^2(\Omega)\underset{-\text{div}}{\longleftarrow}
    L^2(\Omega,\mathbb{R}^3)\underset{\text{rot}}{\longleftarrow} 
    L^2(\Omega,\mathbb{R}^3)\underset{-\text{grad}}{\longleftarrow}
    L^2(\Omega)\;
\end{gather*}
without any boundary conditions. The Laplace-Beltrami operator on forms,
\[
-\Delta = \bigoplus_{p=0}^3(-\Delta_p)= dd^* + d^*d\;,
\]
is seen to correspond to the Euclidean Laplacian with boundary conditions
\begin{equation}\label{sette}\begin{aligned}
{}&{\phi} = 0  &&\qquad\qquad&&(p=0)\;,\\
{}&{\E}_\parallel = 0\;,\quad \dive{\E} = 0 && &&(p=1)\;,\\
{}&{\B}_\perp = 0\;,\quad (\rot {\B})_\parallel = 0 
&& &&(p=2)\;,\\
{}&(\grad \psi)_\perp = 0 && &&(p=3)\;.
 \end{aligned}\end{equation}
Each of the four problems admits a heat kernel expansion,
\begin{equation}
\Tr_{\Lambda_p(\Omega)}\e^{\Delta_pt}\cong \sum_{n=0}^{\infty}
a_n^{(p)}t^{\frac{n-3}{2}}\;,
\label{eq:hke}
\end{equation}
whose coefficients have been computed ($n = 0,\dots,3$) \cite{BBG} or 
can be computed using existing results ($n=4,5$) \cite{BGKV}.
To this end we note that the boundary conditions for $p=1,2$ can be 
formulated equivalently as
\begin{equation}\begin{aligned}
\E_\parallel= 0\;,&\quad \frac{\p \E_\perp}{\p n} - (\tr L)\E_\perp = 0 
&\qquad\qquad&(p=1)\;,\\
\B_\perp = 0\;,&\quad \frac{\p \B_\parallel}{\p n} -  L\B_\parallel = 0 
&\qquad\qquad&(p=2)\;.
\end{aligned}
\label{bdry}
\end{equation}

\noindent
{\bf First approach.} We will compute (\ref{eq:trea}). We observe 
that $-\Delta_{\E}$ is just the 
restriction of $-\Delta_1$ to its invariant subspace
\[
\hc =\left\{\E \in L^2(\Omega,\mathbb{R}^3)\mid  
\dive \E = 0 \right \}= \krn d_1^*\;.
\]
Hence
\[
\Tr'_{\hc}e^{t\Delta_{\E}} = 
\Tr'_{L^2(\Omega,\mathbb{R}^3)}\e^{t\Delta_1}-
\Tr'_{\hc^\perp}e^{t\Delta_1}\;,
\]
where the orthogonal complement of $\hc$ in 
$L^2(\Omega,\mathbb{R}^3)$ is
\[
\hc^\perp = \overline{\Ran d_1} = \Ran d_1 = 
\left\{ \nabla \phi \in L^2(\Omega,\mathbb{R}^3)\mid \phi = 0 
\text{ on } \p\Omega\right\}\;,
\]
($\Ran d$ is closed by the Hodge decomposition, see e.g.~\cite{CFKS, G}). 
By $ d\Delta = \Delta d$, the operators 
$(-\Delta_1)\restriction_{\hc^\perp}$ and $-\Delta_0$ have the same 
spectrum (in fact $\nabla\phi = 0$ implies $\phi=0$ by the boundary 
condition). Thus, using also (\ref{sei}), we find
\begin{align*}
\Tr'_\hc \e^{t\Delta_{\E}} &= \Tr'_\hl \e^{t\Delta_1} - \Tr'_\hs \e^{t\Delta_0}\\
&=\Tr_\hl \e^{t\Delta_1} - \Tr_\hs \e^{t\Delta_0} - (n-1)\;,
\intertext{i.e.,}
a_k &= a_k^{(1)} - a_k^{(0)} \;, \qquad  (k \ne 3)\;,\\
a_3 &= a_3^{(1)} - a_3^{(0)} - n + 1 \;.
\end{align*} 
These relations, together with the values of $a_k^{(p)}$ computed 
in the Appendix, yield the values of the coefficients stated 
in the Theorem~\ref{teo1}. In particular, we will obtain
\[
a_3^{(1)} - a_3^{(0)} = \frac{1}{64}(4\pi)^{-1}\bigl( 3 (\tr L)^2 + 
28\det L \bigr)[\p\Omega]\;.
\]
This matches the stated value of $a_3$ because of
\[
n = \frac{1}{2}\sum_{i=1}^{n}(1 + g_i) + \frac{1}{2}\sum_{i=1}^{n}(1 - g_i)
\]
and of the Gauss-Bonnet theorem,
\begin{equation}
\label{otto}
\frac{1}{2}\sum_{i=1}^{n}(1 - g_i) =  
\frac{1}{2} (4\pi)^{-1}(\det L)[\p\Omega]\;.
\end{equation}
%

\noindent
{\bf Second approach.} We now compute (\ref{eq:treb}). As has been 
noted in the Introduction, eigenmodes of $-\Delta_{\B}$, except for 
zero-modes, satisfy the boundary condition$\  \B_\perp = 0$, and are 
thus eigenmodes of $-\Delta_2$ belonging to its invariant subspace $\hc$, 
cf.~(\ref{dueprimo}, \ref{sette}). The converse is obvious. 
We conclude that
\[
\Tr'_\hc \e^{t\Delta_{\B}} = 
\Tr'_\hl \e^{t\Delta_2} - \Tr'_{\hc^\perp} \e^{t\Delta_2}\;.
\]
Since
\[
\hc = \{ \B \in \hl\mid \dive \B = 0 \} = \krn d_3 \;,
\]
we have
\[
\hc^\perp = \overline{\Ran d_3^*} = \Ran d_3^* = 
\left\{ -\nabla \psi \in L^2(\Omega,\mathbb{R}^3)\mid \psi \in \hs\right\}\;.
\]
Using $d^*\Delta = \Delta d^*$, we see that 
$(-\Delta_2)\restriction_{ \hc^\perp}$ and $-\Delta_3$ have the same 
spectrum, except for a single zero-mode (in fact, $-\nabla\psi = 0$ 
implies $\psi=\const$). We thus find, using (\ref{sei}),
\begin{align*}
\Tr'_\hc \e^{t\Delta_{\B}} &= 
\Tr'_\hl \e^{t\Delta_2} - \Tr'_\hs \e^{t\Delta_3}\\
&=\Tr_\hl \e^{t\Delta_2} - \Tr_\hs \e^{t\Delta_3} - 
\bigl(\sum_{i=1}^{n}g_i\  -1\bigr)\;,
\intertext{i.e.,}
a_k &= a_k^{(2)} - a_k^{(3)} \;,  \qquad (k \ne 3)\;,\\
a_3 &= a_3^{(2)} - a_3^{(3)} -\sum_{i=1}^{n}g_i\  +1 \;.
\end{align*} 
{From} these relations and from the results of the Appendix we again 
recover Theorem~\ref{teo1}. In particular,
\[ 
a_3^{(2)} - a_3^{(3)} = 
\frac{1}{64}(4\pi)^{-1}\bigl( 3(\tr L)^2 - 36\det L \bigr)[\p\Omega]
\]
leads to the claim for $a_3$, because of 
\[
\sum_{i=1}^n g_i = 
\frac{1}{2}\sum_{i=1}^n(1+g_i)-\frac{1}{2}\sum_{i=1}^n(1-g_i)
\]
and of (\ref{otto}).

\section{Application to the Casimir effect}
\label{casi}

For the purpose of this discussion we simply define the Casimir energy by 
the mode summation method, see e.g. \cite{BD2}. In particular, we do not 
address the issue \cite{C} of whether it is the most appropriate 
physically. We shall however observe that the Casimir energy is 
finite -- a conclusion obtained in \cite{BD2}, but 
questioned in \cite{DC}. \\

Consider the cavity $\Omega \subset \mathbb{R}^3$ 
enclosed in a large ball $\Omega_0$. As usual we compare the vacuum 
energy of the electromagnetic field in the domains
$\Omega \cup (\Omega_0 \setminus\overline\Omega)$ with that of the reference 
domain $\Omega_0$. Each eigenmode of either domain contributes a 
zero-point energy $\omega_k/2$, resp. $\omega_k^0/2$.
As a regulator for the eigenfrequencies 
$\omega_k = \lambda_k^{1/2}$, we choose $\e^{-\gamma \lambda_k }$,  
($\gamma > 0$). The corresponding definition of the Casimir energy is 
\[
E_C = \frac{1}{2} \lim_{\Omega_0 \rightarrow \infty}\  \lim_{\gamma \downarrow 0}\left ( \sum_k \lambda_k^{\frac{1}{2}}\e^{-\gamma \lambda_k} \ -\ \sum_k
 (\lambda_k^0)^{\frac{1}{2}}\e^{-\gamma \lambda_k^0} \right )\;.
\]
We shall prove that the limit $\gamma \downarrow 0$ is finite. It will 
also be clear that the subsequent limit $\Omega_0 \rightarrow \infty$ 
exists, though we shall not make the effort to prove that (see 
however e.g. \cite{CFKS}, Section 12.7 for the necessary tools).
Using
$$\lambda_k^{\frac{1}{2}} = -\frac{1}{\sqrt{\pi}}\int_0^{\infty}\ud t\  t^{-\frac{1}{2}} \frac{d}{dt}\e^{-t\lambda_k}$$
and (\ref{eq:asexp}) we find for the regularized sum of the 
eigenfrequencies
\[
\sum_k \lambda_k^{\frac{1}{2}}\e^{-\gamma \lambda_k} \approx 
- \sum_{n=0}^4 \frac{n-3}{2 \sqrt{\pi}}a_n  \int_0^{\delta}\ud t\ 
  t^{-\frac{1}{2}}(t+\gamma)^{\frac{n-5}{2}}
\]
as $\gamma \downarrow 0$. Here $\delta > 0$ is arbitrary, but fixed, 
and ``$\approx$'' means up to terms $O(1)$. Using
\begin{equation*}
\int_0^{\delta}\ud t\  t^{-\frac{1}{2}}(t+\gamma)^{\frac{n-5}{2}} \approx
 \begin{cases} \frac{4}{3}\gamma ^{-2} \qquad \qquad  &(n=0)\;,\\
               \frac{\pi}{2}\gamma^{-\frac{3}{2}} \qquad \qquad &(n=1)\;,\\
               2\gamma^{-1}\qquad \qquad &(n=2)\;,\\
               \pi\gamma^{-\frac{1}{2}}\qquad \qquad &(n=3)\;,\\
               - \log \gamma \qquad \qquad &(n=4)\;,\\
 \end{cases}
\end{equation*}
we find
\begin{equation*}
\sum_k \lambda_k^{\frac{1}{2}}\e^{-\gamma \lambda_k} \approx 
\frac{2}{\sqrt{\pi}} a_0\gamma^{-2}+
\frac{\sqrt{\pi}}{2} a_1 \gamma^{-\frac{3}{2}} + 
\frac{1}{\sqrt{\pi}}a_2 \gamma^{-1} + 
0\cdot a_3\gamma^{-\frac{1}{2}}+ 
\frac{1}{2\sqrt{\pi}}a_4\log\gamma\;.
\end{equation*}
Hence a finite Casimir energy requires (cf. \cite{CVZ}) that 
$a_0,a_1,a_2,a_4$ (but not 
necessarily $a_3$!) agree for $\Omega \cup (\Omega_0 \setminus\overline \Omega)$ 
and for the reference domain $\Omega_0$. This is indeed so for 
$a_0 = 2 (4\pi)^{-\frac{3}{2}}|\Omega_0|$ and for $a_1 = 0$, but also 
for $a_2,\,a_4$ as the contribution from the two sides of $\p\Omega$ 
cancel. The same conclusion is obtained if the regulator 
$\e^{-\gamma \lambda_k}$ is replaced by $\e^{-(\gamma \lambda_k)^{1/2}}$
(see \cite{CVZ}, Eq.~(27)):
\begin{equation*}
\sum_k \lambda_k^{\frac{1}{2}}\e^{-(\gamma \lambda_k)^{1/2}}\approx 
\frac{24}{\sqrt{\pi}} a_0\gamma^{-2}+4 a_1 \gamma^{-\frac{3}{2}} + 
\frac{2}{\sqrt{\pi}}a_2 \gamma^{-1} + 
0\cdot a_3\gamma^{-\frac{1}{2}}+ 
\frac{1}{\sqrt{\pi}}a_4\log\gamma\;.
\end{equation*}
Since no renormalization is necessary, the value of $E_C$ 
agrees with that obtained by means of the zeta function.\\

In the rest of this section we compare our results with those 
of \cite{BD, BD2}. To the extent the comparison is done we will 
find agreement. An important tool there is the mode generating function, 
Eq. (4.5) in \cite{BD},
\begin{equation}\label{nove}
\begin{split}
\Phi(k) &\doteq 
\frac{1}{2} \Tr\left(\frac{-\Delta_{\E}}{-\Delta_{\E}-k^2} + 
\frac{-\Delta_{\B}}{-\Delta_{\B}-k^2}\right)\\
           &\doteq \frac{k^2}{2}\Tr'\Bigl( (-\Delta_{\E}-k^2)^{-1} + 
(-\Delta_{\B}-k^2)^{-1}\Bigr)\;,\qquad 
(k \in \mathbb{C}\setminus \mathbb{R})\;,
\end{split}
\end{equation}
where ``$\doteq$'' means equality ``within addition of some 
polynomial in $k^2$''. Since the resolvents in (\ref{nove}) are not 
trace class, but their squares are, we first consider that replacement. 
Using
$(A+\mu)^{-2} = \int_0^{\infty} \ud t \ t\ \e^{-t(A+\mu)}$
we obtain, as $\mu \rightarrow \infty$,
\begin{equation*}
\frac{1}{2}\Tr'\Bigl( (-\Delta_{\E}+\mu)^{-2} + 
(-\Delta_{\B}+\mu)^{-2}\Bigr) \cong\sum_{n=0}^{\infty}a_n 
\int_{0}^{\infty}\ud t \cdot t^{\frac{n-3}{2}}\e^{-t\mu}
 =\sum_{n=0}^{\infty} \Gamma(\mbox{$\frac{n+1}{2}$})
a_n\mu^{-\frac{n+1}{2}}
\end{equation*}
with coefficients $a_n$ given in Theorem \ref{teo1}.
Integrating w.r.t. $\mu$ we find
\begin{equation*}
\frac{1}{2}\Tr'\Bigl( (-\Delta_{\E}+\mu)^{-1} +
(-\Delta_{\B}+\mu)^{-1}\Bigr)\doteq \sum_{\substack{n=0\\n\ne1}}^\infty
\Gamma(\mbox{$\frac{n-1}{2}$})a_n\mu^{-\frac{n-1}{2}}\ -a_1 \log\mu
\end{equation*}
and hence, with $\mu^{1/2}= -\iu k$,
\begin{equation*}
\Phi(k) \doteq 2 \sqrt{\pi}a_0\iu k^3-\sqrt{\pi}a_1k^2\ln(-k^2) 
+ \iu\sqrt{\pi}a_2k - a_3+O(k^{-1})\;.
\end{equation*}
Upon insertion of the mentioned values for $a_0,\dots,a_3$ this 
agrees with Eq.~(4.40) in \cite{BD}, except for $a_3$ which is there 
replaced by its local part, see (\ref{eq:a3}), 
\begin{equation*}
\tilde{a}_3= 
\frac{1}{64}(4\pi)^{-1}\bigl(3(\tr L)^2-4\det L \bigr)[\p\Omega]
= \frac{1}{64}\int_{\p\Omega}\ud \sigma 
\Bigl(\frac{3}{4}(\kappa_1^2+\kappa_2^2)-\kappa_1\kappa_2 \Bigr)\;,
\end{equation*}
where $\kappa_1,\kappa_2$ are the principal curvatures.
Note however that this discrepancy is implicit in the definition 
of ``$\doteq$''. It is resolved in \cite{BD2} by first considering 
$\delta\Phi (k)$, i.e., the difference of the mode generating functions 
corresponding to the configurations 
$\Omega \cup (\Omega_0 \setminus\overline \Omega)$ and $\Omega_0$. Thus
\begin{equation*}
\delta\Phi (k) = -2\tilde{a}_3 + O(k^{-1}) \;,
\end{equation*}
since the contributions to $a_0,\,a_2$ cancel, and those to 
$\tilde{a}_3$ double the value. Not ambiguous then is ``the number of 
additional modes of finite frequency created by introducing the 
conducting surface $\p \Omega$'':
\begin{equation*}
\mathcal{C} = \psi(0+)-\psi(\infty) \;,
\end{equation*}
where $\psi(y) = \delta\Phi(\iu y)$. For  a connected boundary 
$\p\Omega$ of genus $g$ the value of $\psi(0+)$ has been established 
as $\psi(0+)= -g$ (see \cite{BD2}, Eq.~(5.8)), resulting in
\begin{equation}\label{dieci}
\mathcal{C} = 2\tilde{a}_3 - g\;.
\end{equation}
This result agrees with Theorem \ref{teo1}: the non-local terms 
in (\ref{eq:a3}) take the values 
$-\frac{1}{2}(g-1),\, -\frac{1}{2}g,\, \frac{1}{2}$ for 
$\Omega,\, \Omega_0 \setminus\overline\Omega$ and $\Omega_0$ respectively. Thus,
\[
\delta a_3 = 2\tilde{a}_3 - g\;,
\]
in agreement with (\ref{dieci}).


\appendix
\section{Appendix}

In this appendix we compute the heat kernel coefficients in (\ref{eq:hke}) 
for $p=0,\dots,3$ and
$n=0,\dots,5$ on the basis of Theorems 1 and 4 in \cite{BGKV}. 
We use the same notation, together with $P = \n{n} \otimes \n{n}$ denoting 
the normal projection at the boundary. The vector bundle is 
$V = \Omega \times \mathbb{R}$ for $p = 0,3$, resp. $V = T\Omega$ for 
$p=1,2$, equipped with the Euclidean connection. The decompositions of 
$V  |_{\p\Omega} = V_N \oplus V_D \ni (\phi^N, \phi^D)$ (with 
projections $\Pi_+$, resp. $\Pi_-$) and boundary conditions 
$\phi_{;n}^N + S\phi^N = 0$, resp. $\phi^D = 0$, are specified as 
follows, cf. (\ref{bdry}) and \cite{BGKV}:
\begin{equation}\label{auno}
\begin{aligned}p&=0:  &&\qquad\begin{cases}\Pi_+ = 0\;,\\
                                   \Pi_- = 1\;, \end{cases} \\
               p&=1:  &&\qquad\begin{cases}\Pi_+ = P\;,\qquad S=-L_{aa}P\;,\\
                                   \Pi_- = 1-P\;,\end{cases}\\
               p&=2:  &&\qquad\begin{cases}\Pi_+ = 1- P\;,\qquad S=-L\;,\\
                                   \Pi_- = P\;,\end{cases}\\
               p&=3:  &&\qquad\begin{cases}\Pi_+ = 1\;,\qquad S=0\;,\\
                                   \Pi_- = 0\;.\end{cases} 
\end{aligned} 
\end{equation}
The result is
\begin{align*}a_0^{(p)} &= (4\pi)^{-\frac{3}{2}} c_0^{(p)}|\Omega|\;,\\
              a_1^{(p)} &= \frac{1}{4}(4\pi)^{-1} c_1^{(p)}|\p\Omega|\;,\\
              a_2^{(p)} &= 
       \frac{1}{3}(4\pi)^{-\frac{3}{2}} c_2^{(p)}(\tr L)[\p\Omega]\;,\\      
              a_3^{(p)} &= 
       \frac{1}{384}(4\pi)^{-1}\bigl( c_{31}^{(p)} (\tr L)^2+
        c_{32}^{(p)}(\det L)\bigr)[\p\Omega]\;,\\
              a_4^{(p)} &= \frac{1}{315}(4\pi)^{-\frac{3}{2}}
 \bigl(c_{41}^{(p)}(\tr L)^3+c_{42}^{(p)}\tr L\cdot\det L\bigr)[\p\Omega]\;,\\
              a_5^{(p)} &= \frac{1}{245760}(4\pi)^{-1}
 \bigl( c_{51}^{(p)} (\tr L)^4+c_{52}^{(p)}(\tr L )^2\det L + 
       c_{53}^{(p)}(\det L)^2 + 
       c_{54}^{(p)}\tr L \cdot \nabla ^2 \tr L \bigr)[\p\Omega] \;
\end{align*}
with coefficients given by
\medskip 

\begin{tabular}{c@{\bigg |}cccc}
                   & $\quad p=0\quad$ & $\quad p=1\quad$ & 
                     $\quad p=2\quad$ & $\quad p=3\ $ \\\hline
$ c_0^{(p)}$       &  $ 1$   &   $3$   &  $3$   &   $1$   \\

$ c_1^{(p)}$       &  $-1$   &  $-1$   &   $1$   &   $1$  \\
$ c_2^{(p)}$       &  $1$   &  $-3$   &  $-3$   &   $1$   \\
$ c_{31}^{(p)}$    &  $3$   &  $21$  &  $33$   &  $15$  \\
$ c_{32}^{(p)}$    &  $-20$  &  $148$  & $-220$  &  $-4$  \\
$ c_{41}^{(p)}$    &  $4$   &  $36$   &  $60$   &  $28$   \\
$ c_{42}^{(p)}$    &  $-18$  & $-162$  & $-186$  & $-42$   \\
$ c_{51}^{(p)}$    & $555$ & $5145$ & $8625$ & $4035$ \\
$ c_{52}^{(p)}$    & $- 2840$ & $-27720$ & $-35720$ & $- 10840$ \\
$ c_{53}^{(p)}$    & $2224$ & $29072$ & $29712$ & $2864$ \\
$ c_{54}^{(p)}$    & $120$ & $2520$ & $4680$ & $2280$ \\
\end{tabular}

\bigskip\noindent
These values imply Theorem \ref{teo1}, as explained in its proof.\\

The computation of the table is based on the general result of \cite{BGKV},
which has been applied to (\ref{auno}) using the following identities:
\begin{align*}
\Tr(P_{:a}P_{:b}) &= 2 (L^2)_{ab}\;,\\
\Tr(P_{:a}P_{:a}P_{:b}P_{:b}) &= (L^4)_{aa} + (L^2)_{aa}(L^2)_{bb}\;,\\
\Tr(P_{:a}P_{:b}P_{:a}P_{:b}) &=  2(L^4)_{aa}\;,\\
\Tr(P_{:aa}P_{:bb}) &= 2 L_{ac:a}L_{bc:b} + 4 (L^4)_{aa} + 
  4 (L^2)_{aa}(L^2)_{bb}\;,\\
\Tr(P_{:ab}P_{:ab}) &=  2 L_{ab:c}L_{ab:c} + 6 (L^4)_{aa} + 
  2 (L^2)_{aa}(L^2)_{bb}\;.
\end{align*}
They can be derived by using $\nabla_{\n{e}_a}\n{n}=-L_{ab}\n{e}_b$, so that
\begin{equation*}
P_{:a} = -L_{ac} (\n{e}_c \otimes \n{n} + \n{n} \otimes \n{e}_c)\;,
\end{equation*}
and by assuming without loss that 
$\nabla_{\n{e}_a}\n{e}_b$ has no 
component parallel to $T_p\p\Omega$ at the point $p$ of evaluation, 
i.e., $\nabla_{\n{e}_a}\n{e}_b = L_{ab}\n{n}$. Then
\begin{equation*}
P_{:ab} = -L_{ac:b} (\n{e}_c \otimes \n{n} + \n{n} \otimes \n{e}_c) 
  - 2  (L^2)_{ab}P+(L_{ac}L_{bd} + L_{ad}L_{bc})\n{e}_c\otimes \n{e}_d \;, 
\end{equation*}
from which the above traces follow. In turn they allow the computation 
of similar traces with $P$ replaced by $\chi = \Pi_+ - \Pi_-$, i.e., 
by $\chi = \pm (2P - 1)$ in the cases $p=1,2$. In these two cases we 
also have
\begin{align*}
\Tr S_{:a} &= -L_{bb:a} \;,\\
\Tr S_{:ab} &= -L_{cc:ab} \;,
\end{align*}
and, moreover, for $p=1$,
\begin{align*}
\Tr(S_{:a}S_{:a}) &= L_{bb:a} L_{cc:a} + 2 L_{bb}L_{cc}(L^2)_{aa}\;,\\ 
\Tr(P_{:a}S_{:b})  &= -2 (L^2)_{ab}L_{cc}\;,\\
\Tr(P  S_{:a}S_{:a}) &= L_{bb:a} L_{cc:a} +  L_{bb}L_{cc}(L^2)_{aa}\;,
\end{align*}
resp. for $p=2$,
\begin{align*}\Tr(S_{:a}S_{:a}) &= L_{ab:c} L_{ab:c} + 2(L^4)_{aa}\;,\\ 
\Tr(P_{:a}S_{:a})  &= 2 (L^3)_{aa}\;,\\
\Tr(P  S_{:a}S_{:a}) &= (L^4)_{aa}\;.
\end{align*}
Furthermore, traces of $L^k$, $(k \geq 2)$, were reduced to 
$\tr L,\,\det L$ by means of $L^2 - (\tr L) L + \det L = 0$.
Finally, we used the Codazzi equation, $L_{ab:c} = L_{ac:b}$, as well as
\[
L_{ab:ca}-L_{ab:ac}= L_{aa}(L^2)_{bc} - (L^2)_{aa}L_{bc} \;,
\]
which follows from the Gauss equation.

\bigskip\noindent 
{\bf Acknowledgement.} We thank M. Levitin and G. Scharf for discussions.
The research of D. Hasler was supported in part under the EU-network contract
HPRN-CT-2002-00277.

\end{document}